\documentclass[prb,twocolumn,superscriptaddress,showpacs]{revtex4-1}
\usepackage{graphicx}
\usepackage{siunitx}
\usepackage{bm}
\usepackage{color}
\usepackage{amsmath}
\usepackage{amssymb}
\usepackage{multirow}

\newcommand{\urs}{URu$_2$Si$_2$}

\begin{document}

\title{Resonant x-ray spectroscopy of uranium intermetallics  at the U $M_{4,5}$ edges}

\author{K. O. Kvashnina}
\email{kristina.kvashnina@esrf.fr}
\affiliation{Institute of Resource Ecology, Helmholtz-Zentrum Dresden-Rossendorf, 01314 Dresden, Germany}
\affiliation{The European Synchrotron (ESRF), 38000, Grenoble, France}
\author{H. C. Walker}
\email{helen.c.walker@stfc.ac.uk}
\affiliation{ISIS Neutron and Muon Source, Rutherford Appleton Laboratory, Chilton, Didcot, OX11 0QX, United Kingdom}
\author{N. Magnani}
\affiliation{European Commission, Joint Research Centre, Directorate for Nuclear Safety and Security, Postfach 2340, D-76125 Karlsruhe, Germany}
\author{G. H. Lander}
\affiliation{European Commission, Joint Research Centre, Directorate for Nuclear Safety and Security, Postfach 2340, D-76125 Karlsruhe, Germany}
\author{R. Caciuffo}
\affiliation{European Commission, Joint Research Centre, Directorate for Nuclear Safety and Security, Postfach 2340, D-76125 Karlsruhe, Germany}

\date{\today}
\begin{abstract}
We present resonant x-ray emission spectroscopic (RXES) data from the uranium intermetallics UPd$_3$, USb, USn$_3$ and \urs\, at the U $M_{4,5}$ edges and compare the data to those from the well-localized $5f^2$ semiconductor UO$_2$. The technique is especially sensitive to any oxidation of the surface, and this was found on the USb sample, thus preventing a good comparison with a material known to be $5f^3$. We have found a small energy shift between UO$_2$ and UPd$_3$, both known to have localized $5f^2$ configurations, which we ascribe to the effect of conduction electrons in UPd$_3$. The spectra from UPd$_3$ and \urs\,are similar, strongly suggesting a predominant $5f^2$ configuration for \urs. The valence-band resonant inelastic x-ray scattering (RIXS) provides information on the U $P_3$ transitions (at about $18$~eV) between the U $5f$ and U $6p$ states, as well as transitions of between $3$ and $7$~eV from the valence band into the unoccupied $5f$ states. These transitions are primarily involving mixed ligand states (O $2p$ or Pd, Ru $4d$) and U $5f$ states. Calculations are able to reproduce both these low-energy transitions reasonably well.
\end{abstract}


\maketitle

\section{INTRODUCTION}
The challenge of determining the most probable number of $5f$ electrons in actinide intermetallic compounds is one that has been discussed for the last half century. The fact that U, Np, and Pu can have multiple valence states in chemical compounds introduces an element of uncertainty that does not exist for most intermetallic lanthanide ($4f$) systems, in which the valence state is predominantly Ln$^{3+}$. Direct methods of determining the number of $5f$ electrons are surprisingly rare; one of the oldest methods is by measuring the susceptibility as a function of temperature and extracting from the slope the effective moment. Whereas this gives reasonably unique answers for Pu, for U the effective moments for U$^{3+}$ ($5f^3$) and U$^{4+}$ ($5f^2$) are essentially identical.

For lanthanides, the spectroscopic method of neutron inelastic scattering is able to observe transitions between crystal-field levels in the ground-state J-multiplet (intramultiplet transitions) that can uniquely identify the number of 4f electrons \cite{Turberfield,Birgeneau,Fulde}. When such measurements started in the 1970s there was a surprise that crystal-field transitions in intermetallic actinide compounds were so difficult to observe in comparison with those from lanthanide systems \cite{Wedgwood,Handbook19}. The accepted explanation for this difficulty is that the hybridization of the $5f$ and conduction-electron states broadens the crystal-field transitions so that they are difficult to observe \cite{Hu}. Intermultiplet transitions represent another possible method \cite{Handbook14,Moze}, which is again successful for the lanthanides , but since the energies separating the ground and first-excited states for the actinides are larger than in the lanthanides (greater spin-orbit splitting and also greater crystal-field potential for the actinides), the experiments are that much harder. Again only a few successful studies are reported on UPd$_3$ \cite{Osborn} and on \urs \cite{Park}, and those only for $5f^2$ systems, where the first excited level is $\sim400$~meV, whereas for $5f^3$ configurations these excited levels are expected to be in the range of $550 - 750$~meV \cite{Jones} and have not yet been observed directly.

Even though it is not an intermetallic, the case of UO$_2$ is instructive as it represents a classic system with unquestionably a localized $5f^2$ configuration \cite{Santini}. Crystal-field calculations were first performed in the 1960s \cite{Rahman} but a direct observation was not obtained until the first spallation neutron sources became available \cite{Kern,Amoretti} in the 1980s. The crystal-field potential was then found to be a factor of 3 smaller than proposed in Ref. \onlinecite{Rahman}. Intermultiplet transitions in UO$_2$ were reported by using optical techniques \cite{Schoenes} and are in the energy range expected. However, such optical techniques are much more difficult to apply to intermetallic compounds because of the lack of transparency as well as the observation of multiple phonon modes, and there are very few reports of successful studies.

Of course, from a band-structure perspective, the number of $5f$ electrons around the uranium nucleus in any intermetallic is not necessarily an integer number, and indeed many theoretical studies \cite{Handbook17} have shown that the mean number of $5f$ states in U-intermetallics, as well as uranium metal, is $\sim2.7$. However, we know that the crystal-field potential is important, so what is its effect on these $5f$ states? For example, in the debate on the electronic state of \urs \cite{Mydosh}, in which the material is believed to have no long-range magnetic order at $T = 0$~K, this suggests a singlet ground state, which is possible only in the non-Kramers configuration with an even number of $5f$ states, i.e. $5f^2$ for the uranium ion. The interplay between band states (normally associated with itinerant electron states) and discrete crystal-field levels (normally associated with localized $5f$ states) has, of course, been at the heart of discussions on light actinides, again for half a century.

Synchrotron radiation, and a huge surfeit of spectroscopic techniques that have become available at such sources, should certainly give us new insights into the electronic configurations. The most straightforward are based on absorption spectroscopy, and these were already performed in the 1980s at the most available of absorption edges, that of the $2p\rightarrow6d$ transitions ($L_{2,3}$) \cite{KalkowskiSSC,Bertram} This same group extended the absorption spectroscopy to the $M_{4,5}$ edges (transitions from the $3d$ core states to the partially filled $5f$ states) at the same time \cite{KalkowskiPRB}. These measurements were useful, but limited in their resolution by the large intrinsic core-hole interaction at the different edges. Thus, at the $L$ edges the interaction lifetime results in an intrinsic linewidth of $8 - 10$~eV, and at the $M$ edges to $\sim4$~eV. Since the energy differences between configurations are usually less than these energies, uncertainty is introduced. More recently, resonance X-ray emission spectroscopy (RXES) \cite{Rueff,Vitova,KvashninaJES} has become available at a number of synchrotron facilities. In this technique, the energy of the outgoing fluorescence after the absorption process is analyzed. In this way, the energy resolution in the absorption process may be improved, since the final transitions are from intermediate states with smaller intrinsic linewidths.

Booth et al. \cite{BoothPNA,BoothJES,BoothPRB} have presented RXES data at the $L_3$ edge on a number of actinide intermetallic compounds showing two interesting developments. First, that the edge position (i.e. the absorption peak) can be defined much better with this technique (at the U $L$ edge the resolution is reduced from that given by the core-hole lifetime of $\sim8$~eV to about $4$~eV) and this value, when set against a standard such as the actinide dioxides, seems to be proportional to the density of states at the Fermi level (as measured, for example, by the Sommerfeld coefficient). Second, by analyzing the RXES spectra the curves can be fitted to extract the proportion of contributions from different $5f$-electron configurations.

To add to the discussion about the ground-state configurations of uranium intermetallics, we report in this paper similar experiments to those performed by Booth et al. \cite{BoothPNA,BoothJES,BoothPRB}, but at the uranium $M_{4,5}$ edges. To our knowledge, such measurements have only been reported on UO$_2$ \cite{KvashninaPRL} and other uranium complex systems \cite{KvashninaJES,Butorin2017,Bes,Butorin2016,ButorinPNAS}, so these efforts on U-intermetallics, particularly \urs, should be of interest to those working in this field.

\section{EXPERIMENTAL DETAILS and CALCULATIONS}\label{Sec:ExpDet}
The measurements were performed at beamline ID26 \cite{Gauthier} of the European Synchrotron Radiation Facility (ESRF) in Grenoble. The incident energy was selected using the $(111)$ reflection from a double Si crystal monochromator. Rejection of higher harmonics was achieved by three Si mirrors at angles of $3.0$, $3.5$ and $4.0$~mrad relative to the incident beam. RXES spectra were measured using an X-ray emission spectrometer (XES) \cite{Glatzel,KvashninaJSR}, where the sample, analyzer crystal and silicon drift diode (Ketek detector) were arranged in a vertical Rowland geometry. The full core-to-core RXES data were measured by scanning the incident energy at different emission energies around the $M_\alpha$ and $M_\beta$ lines, near the U $M_5$ and U $M_4$ edges, respectively. Line scans at the maximum of the $M_\alpha$ and $M_\beta$ emission lines are referred to as “high-energy resolution fluorescence detected (HERFD)” absorption spectra. The intensity was normalized to the incident flux.

The emission energy was selected using five spherically bent Si$(220)$ crystal analyzers (with $1$~m bending radius) aligned at $75$~deg. Bragg angle for the measurements at the U $M_4$ edge and using the $(220)$ reflection of Ge analyzers aligned at $78$~deg. Bragg angle for the measurements at the U $M_5$ edge. The paths of the incident and emitted X-rays through air were minimized to avoid losses in intensity due to absorption by air. Combined (incident convoluted with emitted) energy resolutions of $0.4$~eV and $0.3$~eV were obtained at the U $M_4$ and U $M_5$ edges, respectively, as determined by measuring the full width at half maximum (FWHM) of the elastic peaks. A full discussion of the resolution effects in both $L$ and $M$-edge RXES experiments with UO$_2$, and other ionic U compounds is given in Ref.~\onlinecite{KvashninaJES}. The width of the RXES spectrum gives an idea of the energy width (convoluted with the $0.3-0.4$~eV resolution noted above) of the unoccupied $5f$ states above $E_F$.

The valence-band resonant inelastic x-ray scattering (RIXS) data at the U $M_5$ edge have been recorded using the five spherically bent Si crystal analyzers aligned at $65$~deg. Bragg angle and resulted in $1.0$~eV of total energy resolution.

The RXES and RIXS experiments have been performed in identical conditions by placing each U sample in the focus position of the X-ray emission spectrometer.

The data are not corrected for self-absorption effects. The analysis shown in this work is not substantially affected by self absorption, as we are interested in energy positions rather than absolute intensities.

The experiments were performed at room temperature. Since the best resolution we have corresponds to $\sim0.3$~eV (roughly $3400$~K), we do not expect to observe any changes on cooling the sample. The samples were a series of uranium intermetallics, \urs\,(single crystal), UPd$_3$ (single crystal), USb (single crystal), USn$_3$ (solid piece), and a sample of UO$_2$ (pressed pellet). The intermetallic samples were sealed in an argon glove box with a kapton covering of $50$~$\mu$m, with a second encapsulation of $12$~$\mu$m kapton. UO$_2$ was prepared as a pressed pellet and covered by $25$~$\mu$m kapton. Despite these precautions, as we shall see, some oxidization occurred for the USb and USn$_3$ samples.  This is a major difficulty with working at the relatively low-energy beams of $\sim4$~keV as the beam penetration is of the order of $200$~nm at most, so the experiment is sensitive to any near-surface contamination.

Analyses of the RIXS data were performed with the help of theoretical calculations using the FEFF $9.6$ code. FEFF is an \emph{ab initio} multiple-scattering code for calculating the electronic structure and excitation spectra, including local density of states (DOS) \cite{Rehr}. The FEFF code was used to obtain the DOS of the UPd$_3$, UO$_2$, and \urs\, compounds, and these were used as inputs for calculations of the RIXS data to compare with experiment.

The full multiple scattering calculations were performed using a Hedin-Lundqvist self-energy correction in a cluster of $6.0$~\AA\, radius, using the standard routines. Crystal structures reported in the literature were used to generate the input files for the atomic positions.

The RIXS process here has been identified as a convolution of the occupied and unoccupied DOS, taken from FEFF calculations. Such a theoretical description of the RIXS process was discussed in Refs. \onlinecite{Jiminez,KvashninaAC}, and provides a correlation function between filled and empty electronic states. We will show that hybridization of the different molecular orbitals plays an important role, and should be taken into account, while using such a simplified approach for calculations \cite{KvashninaJES,KvashninaAC}.

The quantitative empirical analysis for the $5f$ electron count ($n_f$) was performed using the HERFD spectrum at the U $M_4$ edge of \urs\ by an iterative transformation factor analysis program \cite{RossbergABC,RossbergEST}, which has been successfully applied to the studies of the actinides by the extended X-ray absorption fine structure (EXAFS) technique. In the present paper, the fractions of the $5f^2$ and $5f^3$ configurations in the U $M_4$ edge spectrum of the \urs\,sample have been derived. The analysis shows that by using the linear combination of two components - the spectrum of UPd$_3$ (for the $5f^2$ contribution) and the spectrum of USn$_3$ (for the $5f^3$ contribution), the \urs\,spectrum can be well reproduced – see Sec~\ref{Sec:HERFD} below.

\section{RESULTS}

\begin{figure}
	\begin{center}
		\includegraphics[width=.975\linewidth]{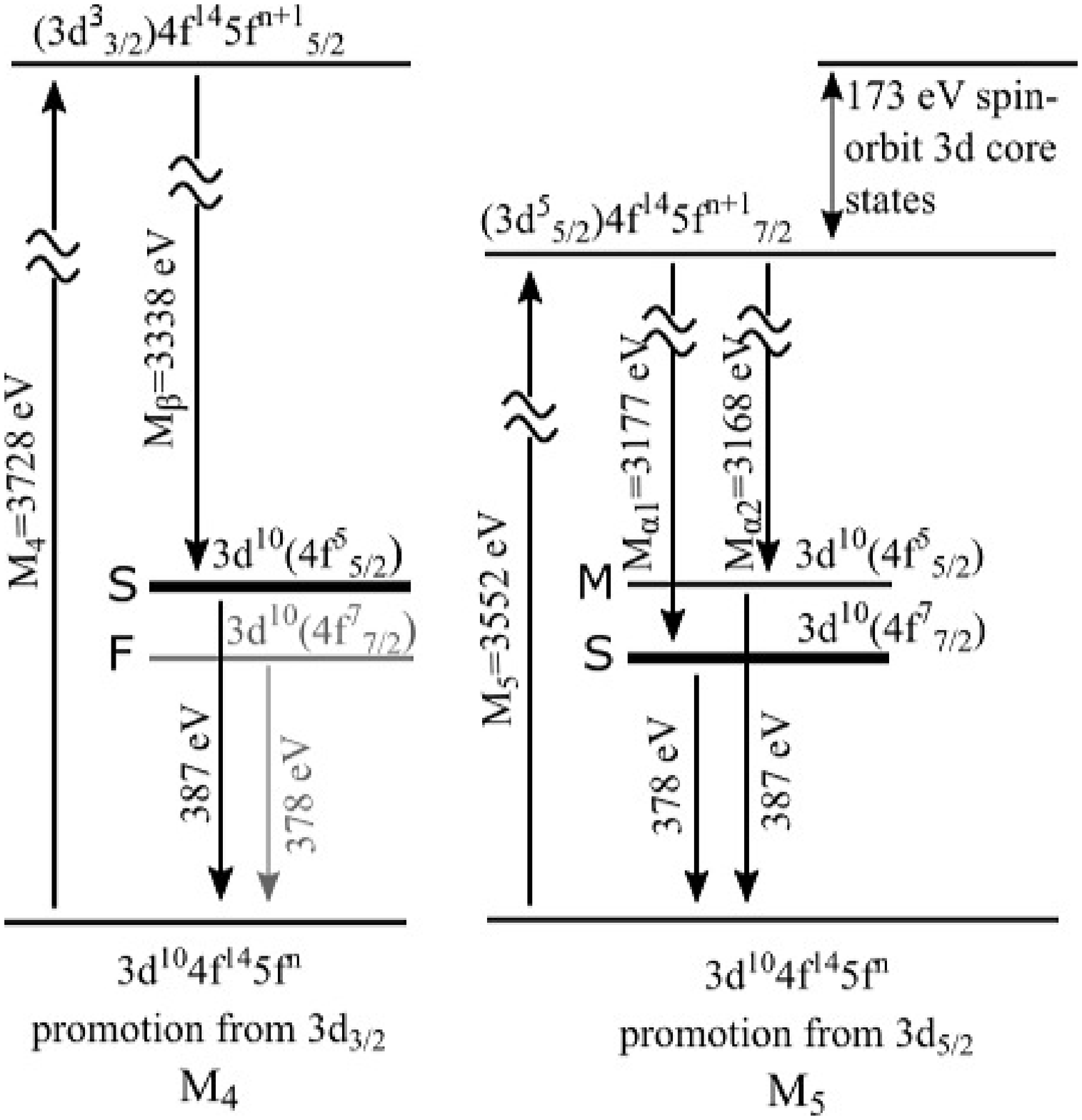}
	\end{center}
    \caption{Spectroscopic scheme for RXES experiments at the actinide $M_{4,5}$ edges. Transition strengths are indicated as \textbf{S} – strong, \textbf{M} – medium, and \textbf{F} – forbidden. The energy values here correspond to the NIST tables \cite{xraytrans} and may not exactly correspond to values in the following figures due to small calibration errors. The diagram takes into account the $10$ electrons of the $3d$ shell being distributed as $3d^4_{3/2}$ and $3d^6_{5/2}$ and the $14$ electrons of the $4f$ shell distributed as $4f^6_{5/2}$ and $4f^8_{7/2}$.}
	\label{Fig:1}
\end{figure}

In X-ray absorption near-edge spectroscopy (XANES), the electron is promoted from the ground state to the first unoccupied state.  The core hole that is created by that process is unstable and is quickly filled by an electron from another level.  The X-ray photons emitted during this process may be measured by XES. Figure~\ref{Fig:1}  shows a schematic representation of the electronic transitions of the XANES and XES processes at the U $M_4$ and $M_5$ edges.

In our RXES experiment at the U $M_{4,5}$ edges we probe the transitions from the ground electron shell – $3d^{10}4f^{14}5f^n$ to the $3d^94f^{14}5f^{(n+1)}$ shell in the U atom and, at the same time record the event when the electrons from the core occupied shells fill the created hole at the ground states  - $3d^94f^{14}5f^{(n+1)}$ to the $3d^{10}4f^{13}5f^{(n+1)}$ . Due to the dipole selection rules ($\Delta J = 0$; $\pm1$) the unoccupied $5f$ electronic levels with $J = 5/2$ and $J = 7/2$ can be reached at the U $M_5$ edge (promotion from the $3d_{5/2}$ state), whereas only the $J = 5/2$ state can be reached at the U $M_4$ edge (promotion from the $3d_{3/2}$ state) (cf. Figure~\ref{Fig:1}). Our paper reports dipole transitions at the $M_{4,5}$ edges. Because of selection rules, directional effects could only be expected in such dipole transitions if the systems had $2$~-fold symmetry or below. Since all the compounds examined have higher symmetry the signal may be accurately considered to have spherical symmetry.

\subsection{High-energy resolution fluorescence data}\label{Sec:HERFD}

\begin{figure}
    \begin{center}
        \includegraphics[width=.9\linewidth, angle=0]{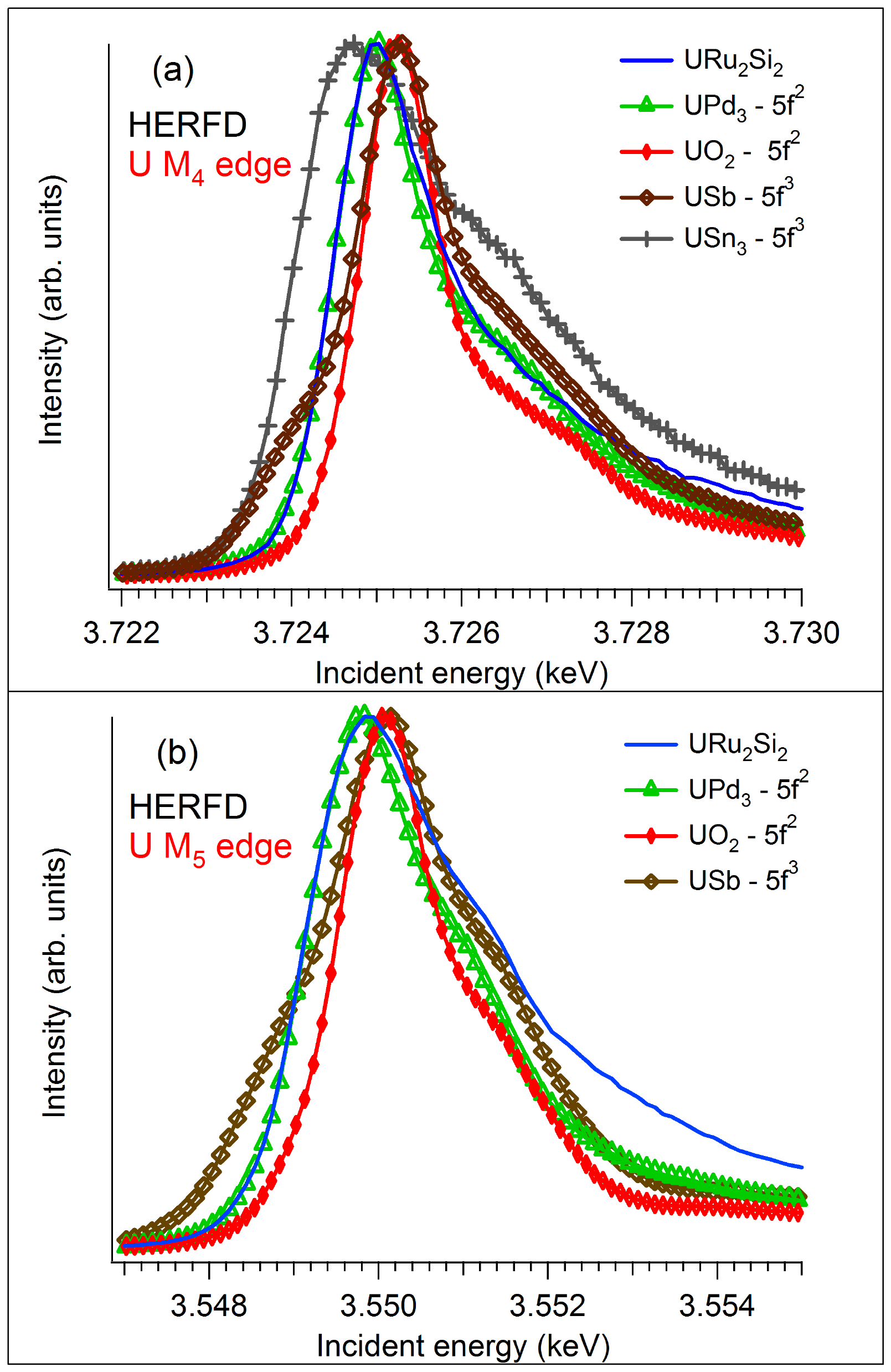}
    \end{center}
    \caption{High-energy resolution (HERFD) data taken at (a) $M_4$ and (b) $M_5$ edges recorded with X-ray emission spectrometer for \urs\, and compared to UO$_2$, UPd$_3$, USb and USn$_3$ (only Fig. 2a) reference systems.}
    \label{Fig:2}
\end{figure}

In Fig.~\ref{Fig:2} we show the HERFD scans taken at the $M_4$ (Fig.~\ref{Fig:2}a) and $M_5$ (Fig.~\ref{Fig:2}b) edges with the emission spectrometer tuned to maximum of the $M_\beta$ and $M_\alpha$ lines, respectively. The same tendency in the shape and position of the main absorption features in HERFD spectra was recorded at both the U $M_4$ and $M_5$ edges, giving confidence in the results. The only difference is the greater broadening of the U $M_5$ HERFD features compared to the U $M_4$ spectra. When comparing these two edges in Figs. 2a and 2b, three factors need to be considered: the core-hole lifetime broadening of the $3d_{3/2}$ ($M_4$ edge) -- $3.54$~eV vs. the $3d_{5/2}$ ($M_5$ edge) level -- $3.94$~eV \cite{xraytrans}; the effects of the interaction of these core holes in the final state of the spectroscopic process with U $5f$ electrons; and the instrumental resolution, which is similar for both experiments.

\begin{figure}
    \begin{center}
        \includegraphics[width=.9\linewidth, angle=0]{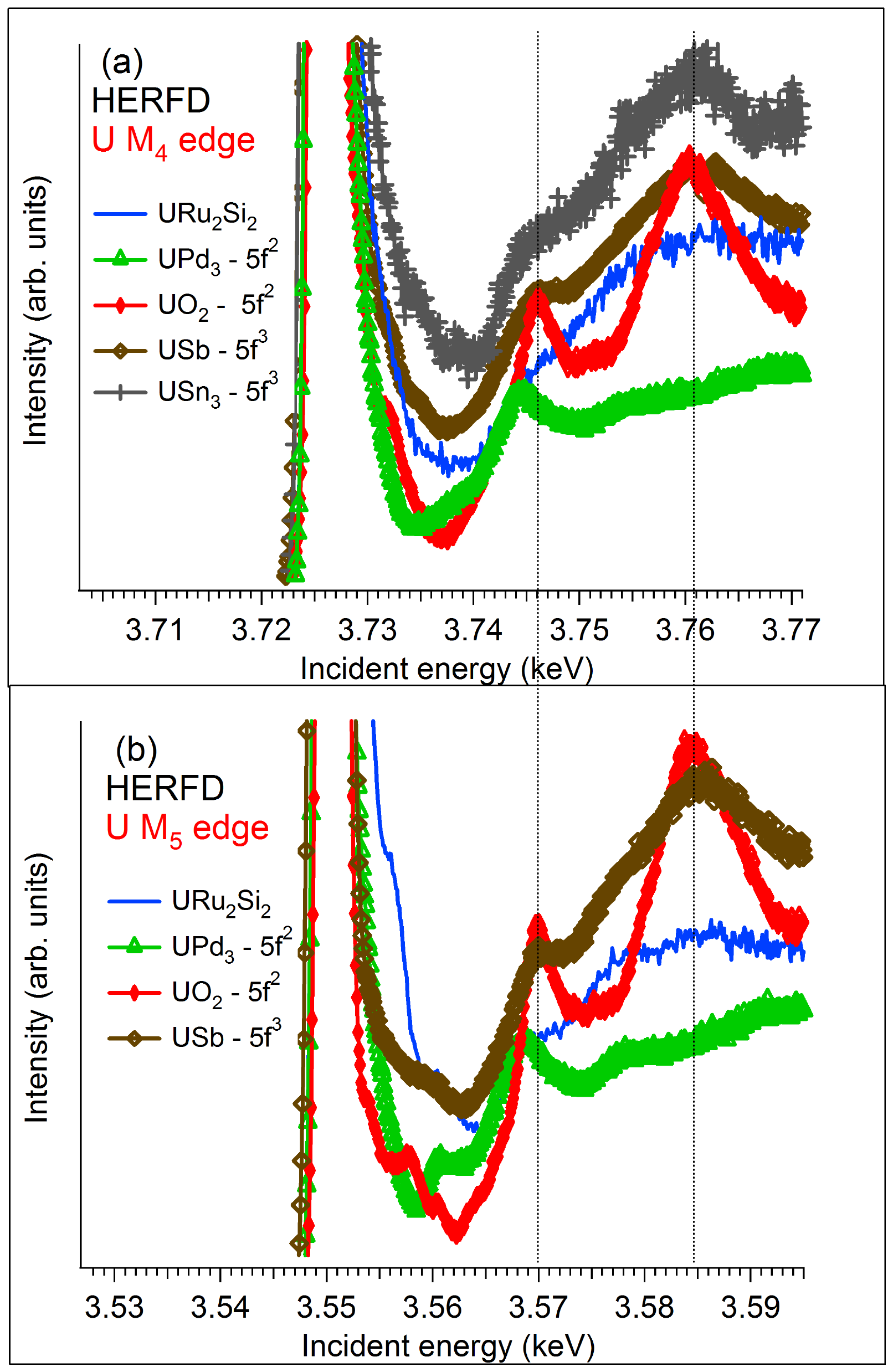}
    \end{center}
    \caption{Post-edge features of the HERFD spectra at the (a) U $M_4$ edge and (b) U $M_5$ edge of UO$_2$, UPd$_3$, USb,  and USn$_3$ (only in Fig.3a.) and \urs\, recorded with the X-ray emission spectrometer set to the maximum of the (a) $M_\beta$ and  (b) $M_\alpha$ emission lines, respectively.}
    \label{Fig:3}
\end{figure}

UPd$_3$ has a $5f^2$ configuration, as judged by the observation of crystal-field levels in neutron scattering and the successful modeling of the ground state \cite{Le}, as well as detailed angular-resolved photoemission experiments \cite{Kawasaki2013} and theory \cite{Yaresko67,Yaresko68}. UO$_2$ also has the same $5f$ configuration \cite{Santini}, but the spectra of these two materials in Fig.~\ref{Fig:2} are not identical. The first noticeable difference between the HERFD spectra of the UO$_2$ and UPd$_3$ is the shift of the white line in the incident energy scale (by $\sim0.2$~eV at the U $M_4$ edge). Secondly, the USb intermetallic system with a nominally pure $5f^3$ ground state configuration shows the strong presence of a $5f^2$ contribution, similar to the UO$_2$ sample. This suggests the oxidation of the surface of the USb sample. The maximum of the HERFD spectrum of the USn$_3$ with $5f^3$ ground state configuration is shifted to the lower incident energy compared to the UPd$_3$ sample (by $\sim0.3$~eV at the U $M_4$ edge). In the case of different oxides we see that the shift from U$^{6+}$ ($5f^0$) to U$^{4+}$ ($5f^2$) is about – $2$~eV at the $M_4$ edge \cite{KvashninaPRL}, so if we assume this is approximately linear we should expect another shift of – $1$~eV for U$^{4+}$ ($5f^2$) to U$^{3+}$ ($5f^3$) in cases when ionic compounds are studied. The shift appears smaller for the intermetallic compounds.

The shift in the peak position in the intermetallic compounds from UPd$_3$ ($5f^2$) to USn$_3$ (which we believe to be close to $5f^3$) is clearly much less than this – $1$~eV, and is closer to – $0.3$~eV. There is some uncertainty in the $5f$ count of USn$_3$. The material has been studied for many years with the initial theory paper suggesting strong hybridization published in 1986 \cite{Koelling}. The Sommerfeld coefficient is $170$~mJ/mole-K$^2$ suggesting it is a heavy-fermion compound \cite{Cornelius}. Neutron scattering finds no sharp crystal field excitations \cite{Loewenhaupt}, unlike UPd$_3$, and more recent nuclear magnetic resonance (NMR) work emphasizes the spin-fluctuation nature of the material \cite{KambePRB,KambePRL}. From these considerations it seems clear that USn$_3$ is probably close to $5f^3$ with $n_f\sim2.7$.

We can assess the oxidation by looking at higher energy to see the EXAFS spectra. The red curves in Fig.~\ref{Fig:3} come from UO$_2$ and show two well-known “peaks” in the EXAFS spectra. The one at $\sim20$~eV from the main emission line is from the nearest U --– O distance, and that at $\sim40$~eV is the signal from the U --– U next nearest neighbor \cite{Conradson}. These are characteristic peaks, and can be used to determine whether the other samples are oxidized or not. Clearly, the near-surface of the USb sample is partially oxidized, and possibly the USn$_3$ to a lesser extent, but both the \urs\,and the UPd$_3$ are not appreciably oxidized.

\begin{figure}
    \begin{center}
        \includegraphics[width=.95\linewidth, angle=0]{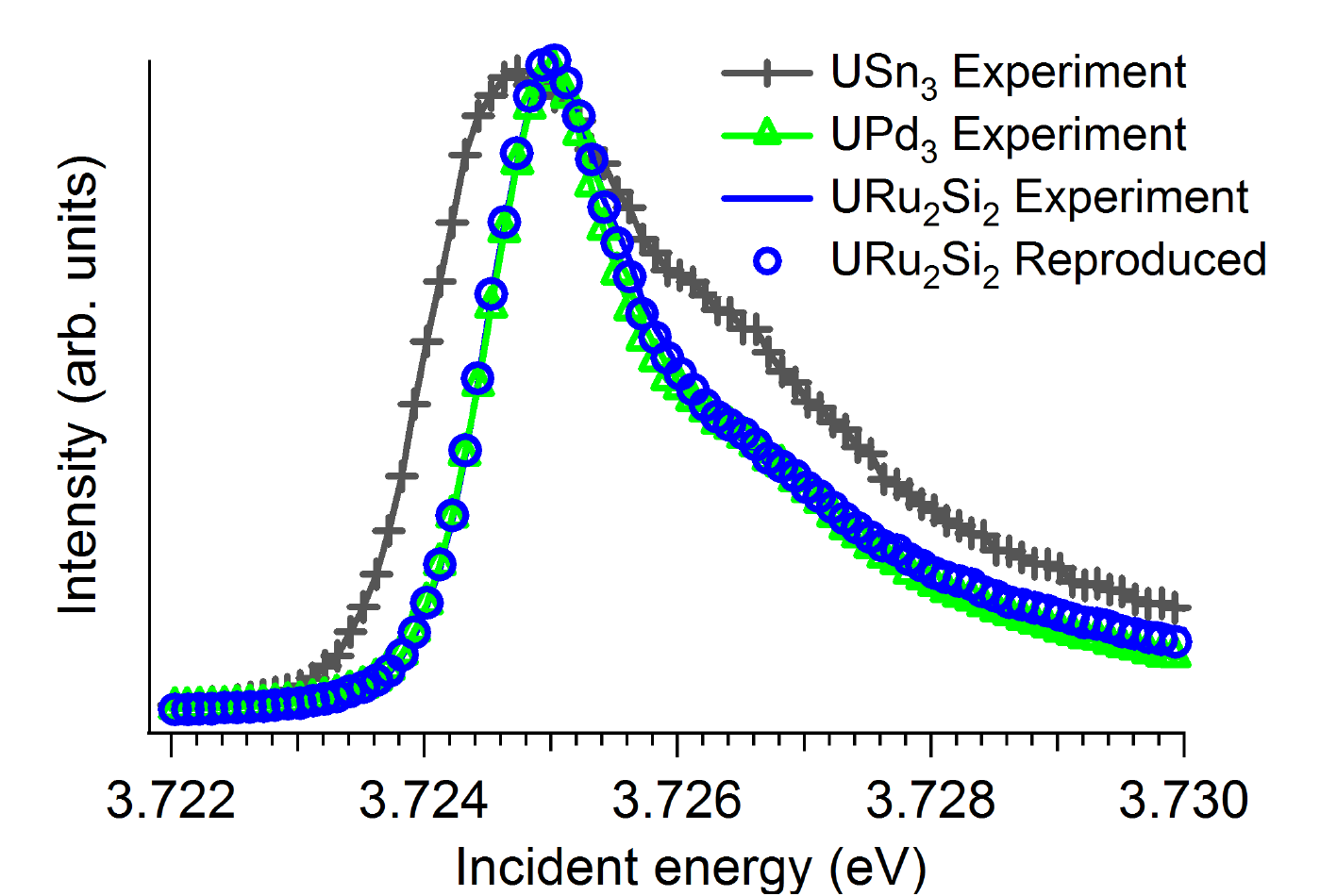}
    \end{center}
    \caption{(a) Experimental (solid lines) and reproduced (open points) U $M_4$ HERFD spectra of \urs\, compared to the UPd$_3$  (taken as $5f^2$) and USn$_3$ (taken as $5f^3$). }
    \label{Fig:4}
\end{figure}

\begin{figure*}
    \begin{center}
        \includegraphics[width=.95\linewidth, angle=0]{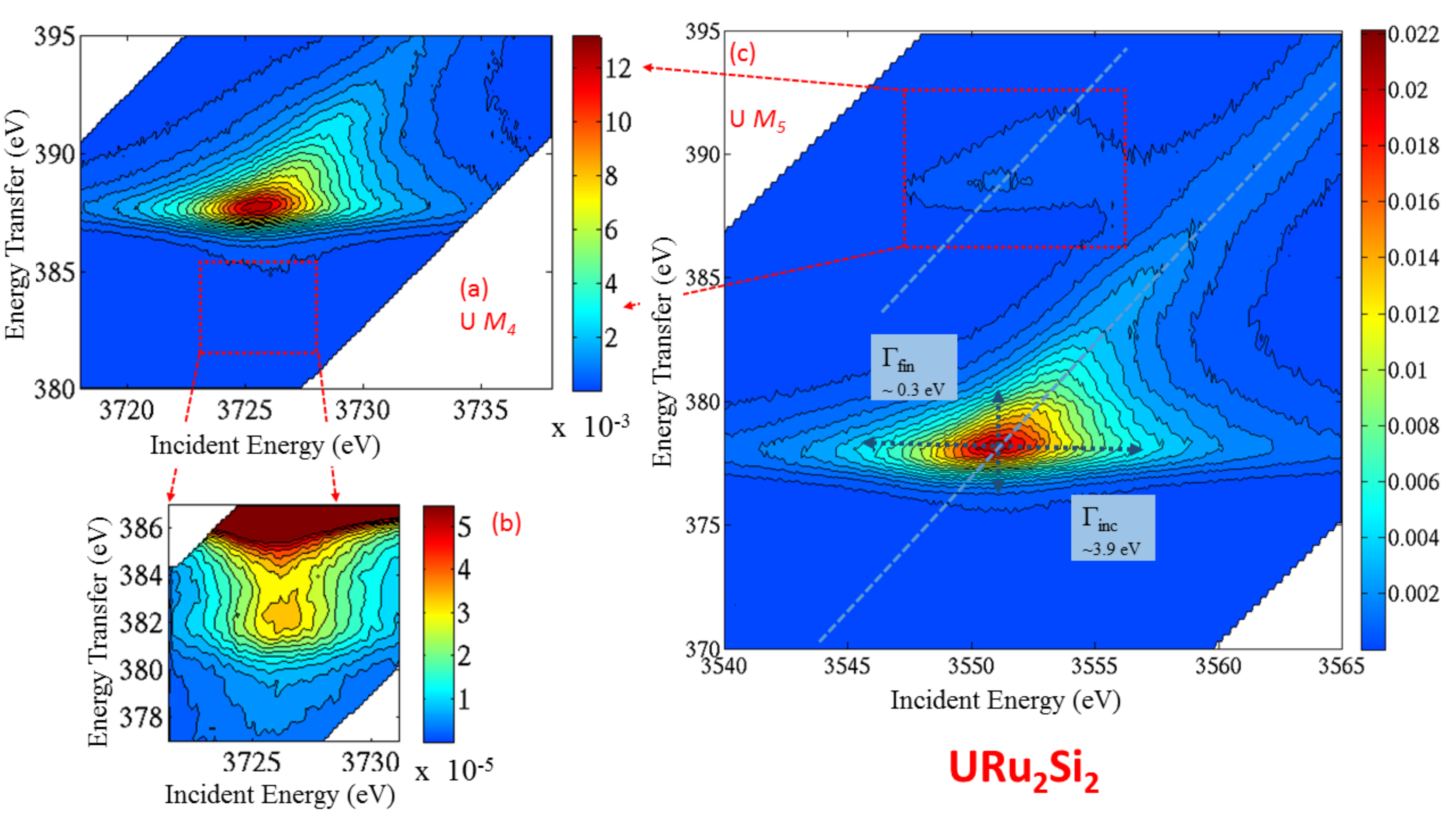}
    \end{center}
    \caption{Full data from RXES experiments at $M_4$ (left) and $M_5$ (right) from \urs.}\label{Fig:5}
\end{figure*}

This effect of oxidation can be also observed in the main edge transitions of HERFD spectra for those samples (Fig.~\ref{Fig:2}). The HERFD spectrum at the U $M_4$ edge of the USb shows the main absorption maximum at $3725$~eV, which is identical to UO$_2$, whereas the USn$_3$ spectrum is broader and peaked at lower energies.

To estimate the possible contribution of $5f^2$ and $5f^3$ configurations in the HERFD spectrum of the \urs, we used the analysis technique described in Sec.~\ref{Sec:ExpDet}. The initial analysis used the HERFD spectrum of UPd$_3$ with a U $5f^2$ ground state configuration and the HERFD spectrum of USn$_3$ with a $5f^3$ configuration as input files. Figure~\ref{Fig:4} shows the comparison of the experimental HERFD spectrum and the reconstructed one for \urs\ by the program and compared to the HERFD spectra of the UPd$_3$ and USn$_3$ reference systems. The results show very little of $5f^3$ is needed, but in view of the possibility that the USn$_3$ spectra are also slightly contaminated by oxide we prefer to increase the error bar, finding $n_f = 2.05 \pm 0.10$.

We can now make a few preliminary conclusions.
\begin{enumerate}
\item The UO$_2$ does not have the peak in the absorption spectrum at the same place as that of UPd$_3$. Since the latter is well characterized as a $5f^2$ system this is perhaps surprising, but one has to remember that the intermetallic systems possess conduction electrons, whereas UO$_2$, which is also $5f^2$, does not. This suggests that taking UO$_2$ as a ``standard'' reference system for the localized $5f^2$ configuration in U intermetallics is inappropriate.
\item The \urs\,spectra at both the $M_{4,5}$ edges fall exactly at the same place as that of UPd$_3$. Since UPd$_3$ is a $5f^2$ system, this strongly suggests the ground state of \urs\,is also close to $5f^2$.
\item Although both the USb and USn$_3$ spectra are not clean (due to the oxidation) there is evidence of intensity at lower energy, as would be expected for $5f^3$. However, the shift from $5f^2$ to $5f^3$ appears considerably less ($0.3$~eV) than found in the oxide systems (assuming some linearity in the oxide systems since for uranium no U$^{3+}$ state exists). Signals from all higher oxidation states (U$^{4+}$ and above) would fall at higher incident energies \cite{KvashninaPRL,Butorin2017,Bes,Butorin2016,ButorinPNAS}.
\end{enumerate}

\begin{figure}
	\begin{center}
		\includegraphics[width=.9\linewidth, angle=0]{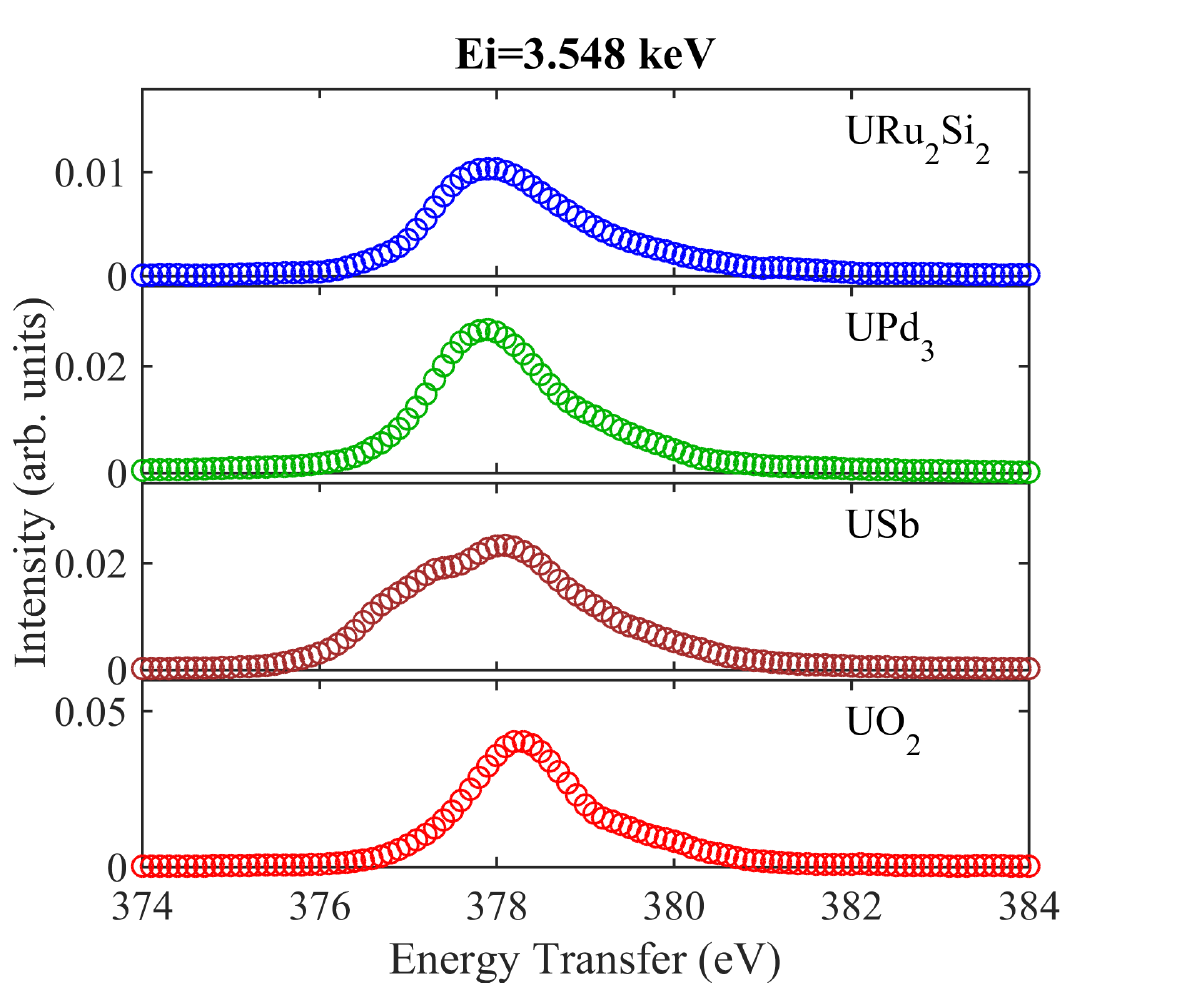}
		\includegraphics[width=.9\linewidth, angle=0]{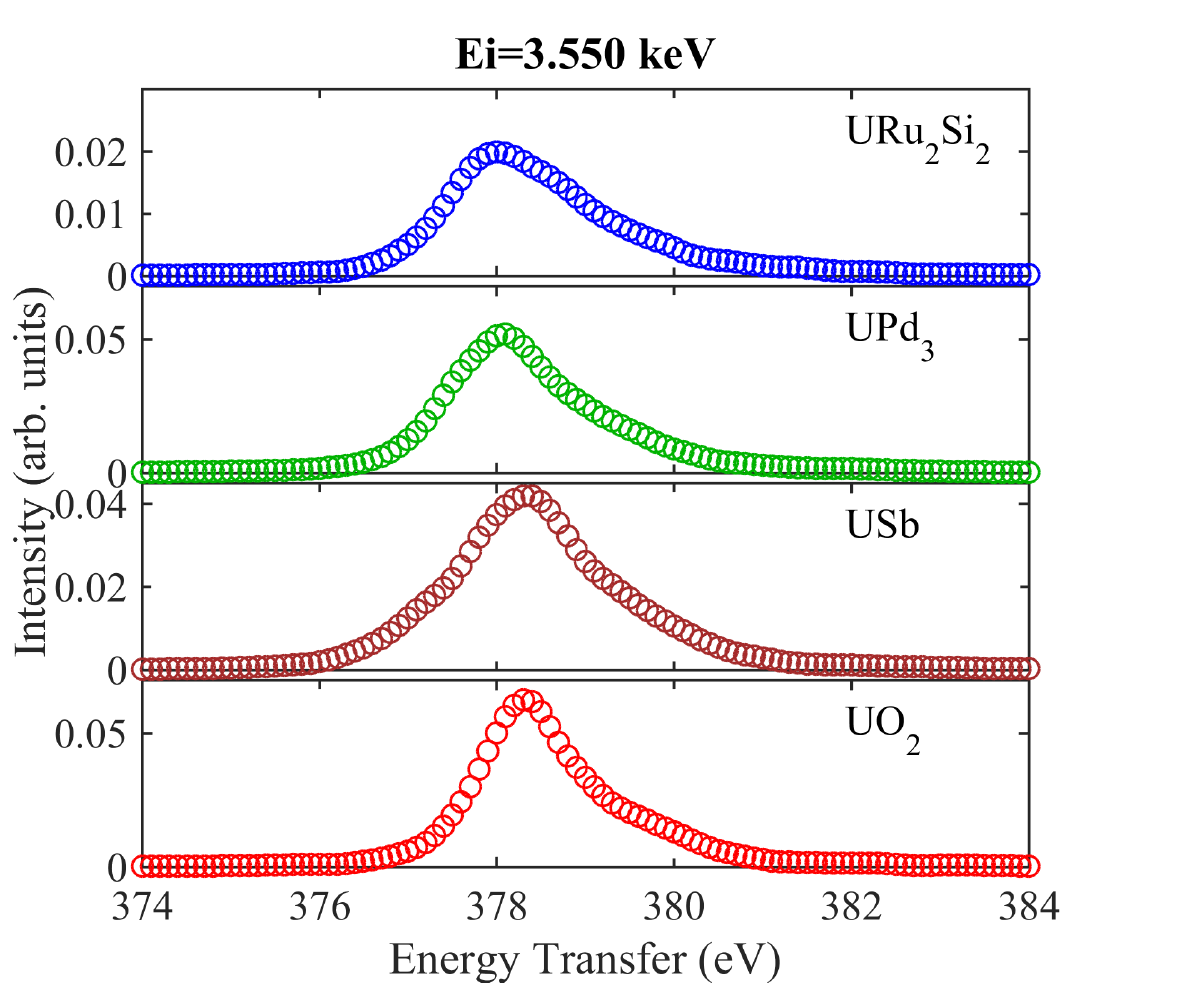}
	\end{center}
	\caption{Intensity, normalized to the incident beam intensity, as a function of energy transfer for fixed incident energies just below the $M_5$ edge for $4$ compounds. }\label{Fig:6}
	\label{}
\end{figure}

\subsection{Resonant x-ray emission data}

The experimental core-to-core RXES maps of the incident photon energies at the U $M_4$ and U $M_5$ edges of \urs\,are shown in Figure~\ref{Fig:5}. Such maps are standard for these experiments and are shown in Ref. \onlinecite{BoothPNA,BoothJES,KvashninaJES,KvashninaPRL,Vitova}. All spectra we have taken are similar to these data. Fig.~\ref{Fig:2} corresponds to data taken at the maximum of $M_\alpha$ and $M_\beta$ emission lines, marked as two dashed  lines along the diagonal in the RXES plane Fig.~\ref{Fig:5}. The dashed arrows through the RXES plane at the U $M_5$ edge indicate the life-time broadening of the absorption process ($\Gamma_\mathrm{inc}\sim3.9$~eV) and the emission process ($\Gamma_\mathrm{fin}\sim0.3$~eV). This broadening is responsible for the shape of the RXES spectra that are extended more in the incident energy direction (horizontal scale) in comparison to the vertical scale.

The spectral intensities extending along two diagonal directions in the RXES plane correspond to the $4f_{5/2}$ and $4f_{7/2}$ final states, i.e. the $M_\alpha$ and $M_\beta$ emission lines, respectively (cf. Fig.1). The energy separation between the two lines is thus the $4f$ spin-orbit interaction ($\sim9$~eV). The strengths of the two final states are clearly observed from the color bar on the right-hand side of the Figure~\ref{Fig:5}. At the $M_5$ edge, the intensity of the $4f_{7/2}$ final state is higher than the one detected for the $4f_{5/2}$ final state. The same final state $4f_{5/2}$ is detected for the core-to-core RXES process at the U $M_4$ edge. The only difference is that core-to-core RXES at the U $M_4$ edge has revealed an additional feature that has not been previously reported. This is the feature in the insert below the $M_4$ spectra. Normally, one might think this is some leakage at the forbidden peak at the $M_4$ (see Fig.~\ref{Fig:1}), but rather than being at an emission energy of $378$~eV, it is at $382\pm1$~eV, which is closer to the main emission line. The energy difference between the $4f$ core states $4f_{5/2}$ and $4f_{7/2}$ is known from photoemission experiments \cite{Bonnelle,Fujimori} to be $\sim9$~eV, and is reflected in the difference observed in the medium and strong $M_5$ lines in Fig.~\ref{Fig:5}. We have, at present, no explanation for this feature below the $M_4$ edge. However, we note that the feature was observed from all samples examined in this study and the strength of this extra feature, as compared with the main line for the $M_4$ incident energy, is $\sim1\%$ for all materials.

\subsection{Cuts at constant incident energy}

\begin{figure*}
    \begin{center}
        \includegraphics[width=.9\linewidth, angle=0]{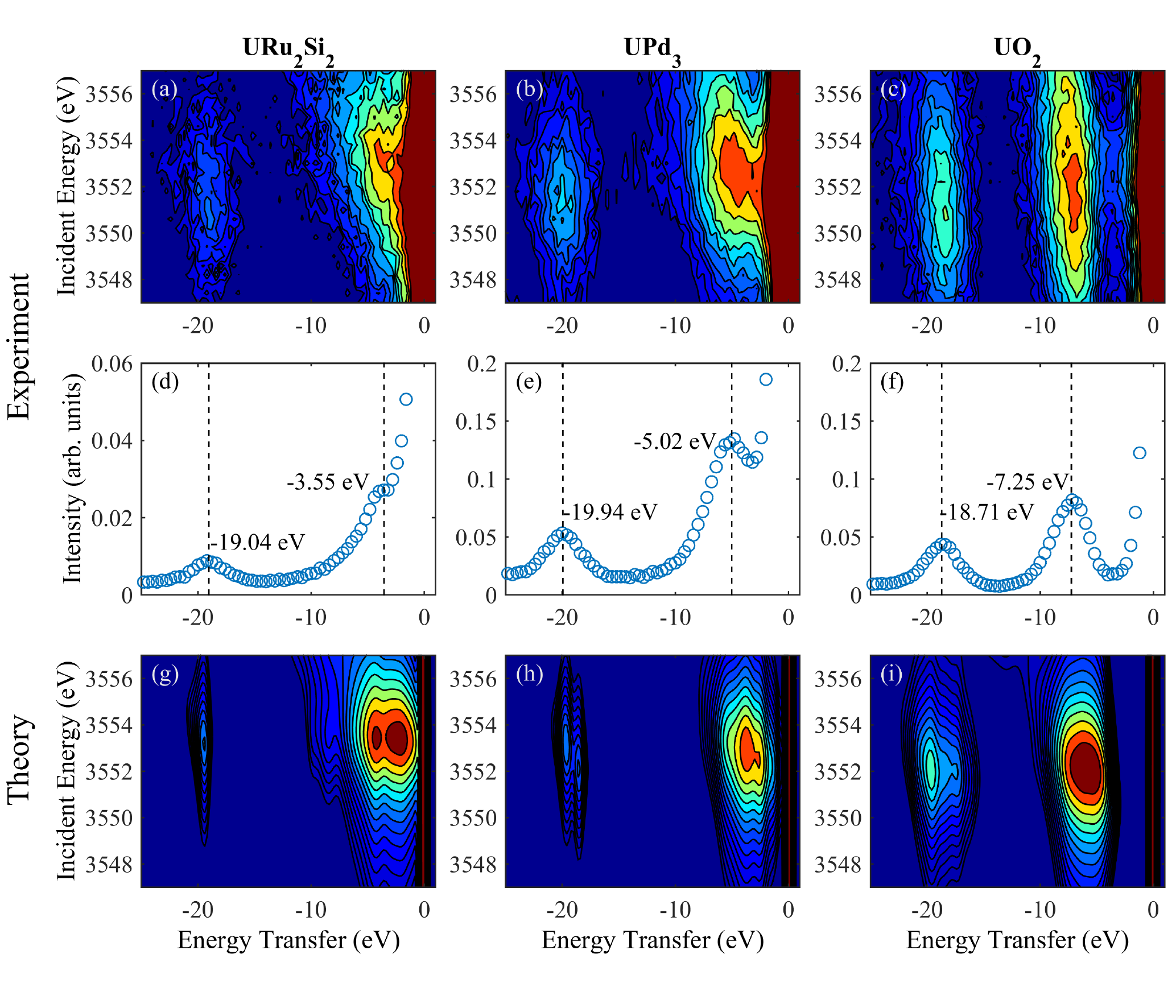}
        \caption{Valence-band spectra of \urs, UPd$_3$, and UO$_2$ around the incident energy of the U $M_5$ edge. The experimental energy resolution is $\sim1$~eV, see Sec. II. The plots (a), (b), and (c) are the experimental data. The plots (d), (e), and (f) are vertical integrations of the experimental data (i. e. as a function of energy transfer). The lower plots (g), (h) and (i) are calculations of the same quantities as discussed in the text and represented by the density-of-states shown in Fig.~\ref{Fig:8}.}\label{Fig:7}
    \end{center}
\end{figure*}

Vertical cuts at fixed incident energies through the RXES data (Fig.~\ref{Fig:5}), and then plotting the intensity as a function of the energy transfer, as shown in Fig.~\ref{Fig:6}, is also a useful way to present the data.  Booth et al. \cite{BoothPNA,BoothJES,BoothPRB} have used these types of cuts in their analysis to obtain the $5f$ count in a number of different actinide compounds at the U $L_3$ edge.

The long tail of the RXES distribution (Fig.~\ref{Fig:5}) will produce an asymmetry in the relevant cut as long as the energy is at or above the resonant energy. This effect is due to transitions into the continuum. To avoid this asymmetrical shape we show only curves taken with the incident energy less than the resonant $M_5$ energy of $3552$~eV.

We have chosen not to analyze these data with Gaussian curves, as done by Booth et al. \cite{BoothPNA,BoothJES,BoothPRB} as the RXES is a two-step process and involves transitions from the ground to the excited states and from the excited to the final states.  Additionally, the conclusions are not substantially different from those given after analyzing Fig.~\ref{Fig:2}. The $5f^2$ profiles of UO$_2$ are shifted to slightly higher emission energy ($378.2$~eV) than those of UPd$_3$ and \urs, which are close to $378.0$~eV.

The spectra for UO$_2$, UPd$_3$, and \urs\,are essentially single functions, at least neglecting some continuum scattering on the high-energy transfer side. On the other hand, USb clearly shows a double peak, with intensity on the low-energy side corresponding to the $5f^3$ contribution.

\subsection{Valence band resonant inelastic X-ray scattering (RIXS)}\label{Sec:RIXS}

We show in Fig.~\ref{Fig:7} data from valence band RIXS taken from three samples. The first noticeable difference between the core-to-core (RXES) and valence-band RIXS data concerns the dispersion of the features. The core-to-core RXES is extended in incident energy and final state direction as discussed previously (cf. Fig.~\ref{Fig:5}). A similar effect would be observed for the valence band RIXS if the data were recorded in an extended energy range. Unfortunately, the valence-band RIXS measurements are time consuming (around $12$~h per sample for the data reported in Figure~\ref{Fig:7}(a)) and we have restricted the recorded incident energy range near the maximum of the absorption edge.

There are two contributions clearly observed in these spectra. The highest energy features at some – $18$~eV are associated with the transitions between the U $5f$ states and the U $6p_{3/2}$ shell\cite{Butorin2000,KvashninaJES}. The process involves first an initial excitation from the $3d$ core state to the unfilled $5f$ state, and then the core hole in the $3d_{5/2}$ core state is filled by an electron from the filled U $6p_{3/2}$ state, with a decay energy (of $\sim18$~eV) back to the ground state.

The tabulated \cite{xraytrans} binding energy for this U $P_3$ ($6p_{3/2}$) transition is $\sim16.8$~eV. Since we are not aware of any calculations for these transitions in different materials, we cannot compare the small changes observed with values available in the literature. However, we definitely observed the slight variation of the X-ray emission energy between different U intermetallic systems of the order of $1$~eV.

The lowest energy feature is a transition from the $3d$ core state to the unoccupied $5f$’s and then the core-hole is filled with an electron from the valence band, with a decay back to the ground state. 

\begin{figure}
    \begin{center}
        \includegraphics[width=0.85\linewidth,angle=0]{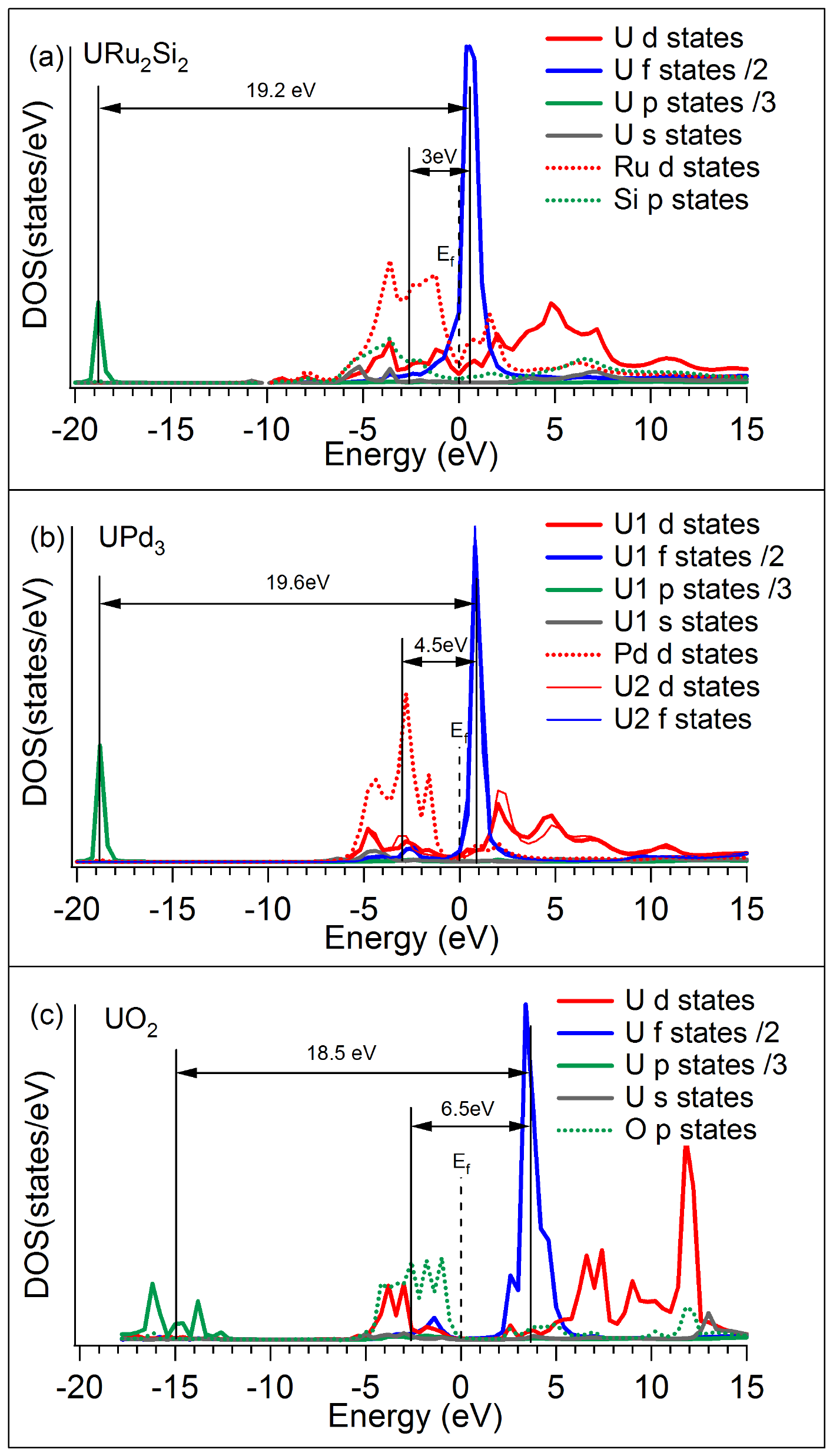}
    \end{center}
    \caption{Partial density of states of (a) \urs, (b) UPd$_3$, and (c) UO$_2$ obtained from FEFF calculations. The most intense peaks are reduced for clarity (the ratio is indicated on the right side of each figure). Fermi level (dashed line) is situated at $0$~eV. }\label{Fig:8}
\end{figure}

To shed more light on the value of these transitions in Fig.~\ref{Fig:7}, we performed RIXS theoretical calculations (as discussed in Sec.~\ref{Sec:ExpDet}) by inserting the partial density of states (DOS), particularly the U $5f$ states and ligand O $2p$ or Ru, Pd $4d$ states, into the Kramers-Heisenberg equation. The partial DOS’s have been calculated for the different materials by the FEFF program and are shown in Figure~\ref{Fig:8}.

Of course, the FEFF codes are not as sophisticated as state-of-the-art treatments of the $5f$ – electron behavior, but they do help us to understand the individual transitions, and, as we shall see, the DOS’s are in reasonable agreement with more advanced calculations. In some cases the level of the Fermi energy ($E_F$) is not correctly obtained from FEFF calculations, and we have shifted the value of $E_F$ to agree with more advanced calculations, for example, Ref. \onlinecite{Yaresko67} for UPd$_3$ and Refs. \onlinecite{Yaresko68,Oppeneer} for \urs. This does not, of course, affect the values of the transitions from the occupied valence band states (i.e. $p$, $d$, or $f$ states) and the unoccupied $5f$ states, which is what is measured and shown in Fig.~\ref{Fig:7}(a) to \ref{Fig:7}(f).

We shall start by discussing the case of UO$_2$, where there have been numerous experiments and theory. Previous experiments clearly place the $5f$ band some $1-2$~eV below $E_F$, and the oxygen $2p$ bands a further $2-3$~eV below this \cite{Cox}, and the BIS experiments \cite{Yu} place the unoccupied $5f$ states some $5$~eV above $E_F$.  This implies that the O $2p$ and U $5f$ energy gap is $\sim7$~eV, and this is precisely what is observed in Figs.~\ref{Fig:7}(c) and \ref{Fig:7}(f). The FEFF calculations (Fig.~\ref{Fig:8}(c)) get a slightly smaller gap for that transition ($6.5$~eV), but as shown in the review \cite{Wen}, there are many calculations with this transition varying from $5-10$~eV. 

In contrast to UO$_2$, for \urs\,and UPd$_3$ there is no clear separation between elastic and inelastic scattering profiles. Theoretical FEFF calculations for \urs\,(Fig.~\ref{Fig:8}(a)) show that the occupied states are dominated by the Ru $4d$ electron bands with a mixture of Si $p$ states and U $6d$ states and distributed over the region $\sim6$~eV below $E_F$. The DOS reported in Figure~\ref{Fig:8}(a) for \urs\,shows a clear hybridization between U $5f$ and mostly Ru $4d$ states. The center of mass of the Ru $4d$ electron band is found to be at $–2.5$~eV below Fermi level. Moreover, the difference between the center of mass distribution of occupied Ru $4d$ states and unoccupied U $5f$ states is found to be $\sim3$~eV, which has to be compared to the $\sim3.5$~eV observed experimentally in RIXS data (Fig.~\ref{Fig:7}(a) and \ref{Fig:7}(d)). These quantities are in agreement with the density of states given in Ref. \onlinecite{Oppeneer}. Additionally the difference between the occupied U $6p$ and unoccupied U $5f$ states is $\sim19.2$~eV, which can be compared to the $19$~eV observed experimentally (Fig.~\ref{Fig:7}(d)). These results are summarized in Table I.

\begin{table}
    \caption{\label{Tab:FEFF} Experimental and calculated (with FEFF program described in text) values (in eV) for transitions between (1) Uranium $3d_{5/2}-6p_{3/2}$ states and (2) valence-band states and unoccupied $5f$ states.}
    \begin{tabular}{c|ccc}
        \hline\hline
                                     & \urs           & UPd$_3$        & UO$_2$ \\
        \hline
        Exp: U $3d_{5/2}-6p_{3/2}$ & $19.04\pm0.50$ & $19.94\pm0.50$ & $18.71\pm0.50$\\
        Theory: FEFF                 & $19.2$         & $19.6$         & $18.5$\\
        \hline
        Exp: valence band – $5f$     & $3.55\pm0.50$  & $5.02\pm0.40$  & $7.25\pm0.25$\\
        Theory: FEFF                 & $3.0$          & $4.5$          & $6.5$\\
        \hline\hline
    \end{tabular}
\end{table}

The investigations of the theoretical DOS for UPd$_3$ also show good agreement with theoretical results reported in the literature \cite{Yaresko67}. There are two inequivalent U positions in the TiNi$_3$ structure and we show in Figure~\ref{Fig:8}(b) the partial DOS for both U atoms. Similar to the case of \urs, $E_F$ has been shifted to the value reported \cite{Yaresko67}. To compare with \urs, $5f$ electron states in UPd$_3$ are not found at $E_F$ but are about $1$~eV below \cite{Beaux}. The strong hybridization by Pd $4d$ with the occupied $5f$ states is visible with the transition energy to the unoccupied $5f$ states being $\sim4.5$~eV, versus $\sim5.0$~eV observed (Fig.~\ref{Fig:7}(e)) experimentally.  The separation between U $5f$ unoccupied states and U $6p$ occupied states is $\sim19.6$~eV (versus $19.9$~eV observed experimentally).

\section{Discussion and conclusions}
\subsection{Overview of spectroscopy}

Despite a number of chemical systems containing uranium being examined by the RXES technique at the $M_{4,5}$ edges \cite{KvashninaPRL,KvashninaJES,Butorin2017,Bes,Butorin2016,ButorinPNAS,Butorin2000}, we believe this is the first detailed report of such spectroscopy of U-intermetallic compounds, apart from a brief summary \cite{Gumeniuk}. The observation that the USb sample had a partially oxidized near-surface region illustrates one of the cautionary tales of this endeavor. Notice that our results for both $M_4$ and $M_5$ edges are consistent with one another, giving confidence in the results.

UPd$_3$ and UO$_2$ are both well-localized $5f^2$ systems: UO$_2$ is a semi-conductor with a $\sim2$~eV bandgap \cite{Santini}, whereas UPd$_3$ is a $5f^2$ localized configuration, probably with $2-3$ electrons in the $6d7s$  conduction band \cite{Osborn,Jiminez,KvashninaAC,Fujimori,Beaux}. That the presence of a conduction band should provoke a difference of $\sim0.2$~eV when both configurations are $5f^2$ in the peak of the $M_{4,5}$ spectra between the two materials is perhaps not surprising, but shows the importance of choosing a standard against which other materials can be calibrated. Both USn$_3$ and, to a lesser extent because of the oxidation, USb have spectral weight at lower incident energies (Fig.~\ref{Fig:2}), which point to a $5f^3$ component in their ground states. This is anticipated for both materials. However, the magnitude of this shift appears to be only about – $0.3$~eV, which is far less than the $\sim1$~eV suggested for insulating oxides between valence states. 

\subsection{Differences between RXES at $L_{2,3}$ and $M_{4,5}$ edges}

Most absorption studies have been performed at the $L_{2,3}$ edges of the actinides, particularly the $L_3$ edge for uranium, which is at $17.17$~keV. It was only natural that the spectroscopic studies using the RXES technique on actinide intermetallics should start with the $L_3$ edge \cite{BoothPNA}. These energies also have the advantage that the beam penetration is several microns, so near-surface effects are of little concern, and X-ray beams of such energies are not attenuated appreciably in air. However, the primary transition is to promote a $2p$ core electron to the partially filled $6d$ valence shell. The transitions are illustrated in Fig. 1A of Ref. \onlinecite{BoothPNA}. The intermediate state, as shown in this figure, involves a hole in the $3d$ core shell, with the emission to the ground state then filling the $2p$ core hole. In this process the $5f$ states are spectators, i.e. they do not play a direct role.  The question is whether the character of the excited states is transmitted directly to the intermediate states?

However, the RXES data at the U $L_{2,3}$ edges give important information on the position of the $3d$ level in different intermetallic systems. The results reported in Ref.~\cite{BoothPRB} show that the maximum of the U $L_{\alpha1}$ ($3d-2p$ transition) at excitation energies above the absorption edge is identical for all investigated intermetallics and UO$_2$. We observe a similar behavior of the U $M_\alpha$ and $M_\beta$ emission lines, indicating that the energy position of the U $4f$ level is identical for all intermetallic systems and UO$_2$. These emission lines are situated $380-390$~eV below $E_F$. We found a difference in the energy position of the U $6p$ level (about $18$~eV below $E_F$) between UPd$_3$, \urs\ and UO$_2$, which has been discussed in section~\ref{Sec:RIXS}.

\subsection{Results for \urs}

A main interest of our experiments is, of course, in the material \urs, which has been much studied since its discovery in the 1980s \cite{Mydosh} and is still controversial. Given that the spectra of \urs\,are almost identical to those of UPd$_3$, suggests that \urs\,is predominantly of $5f^2$ character. This is in agreement with many other spectroscopic techniques using both neutrons \cite{Park} and X-ray techniques such as those using soft resonant X-rays at the U $O_{4,5}$ edge by Wray et al., \cite{Wray}, as well as the most recent non-resonant inelastic X-ray scattering experiments also at the $O_{4,5}$ edges \cite{Sundermann}. These latter experiments are able to go further and even suggest the crystal-field ground state. A similar ground state based on $5f^2$ is suggested by the polarized neutron study of the induced magnetic form factor \cite{Ressouche}. We estimate our error bar on the number of $5f$ electrons, $n_f=2.05\pm0.10$, on the basis that for $\sim5\%$ of $n_f = 3$, we would start to observe intensity in the HERFD spectra at lower energies.

A recent study by Booth et al. \cite{BoothPRB} has given a value for \urs\,of $n_f = 2.87\pm0.08$. As we have discussed above, these measurements use the $L_3$ edge where the primary information is about the $6d$ valence band. If there is strong hybridization between the $5f$ and conduction electrons (mainly $6d$), then it might not be surprising that the experiments at the $L_3$ edge find a larger number for the effective $n_f$.

Notice that we have not stated whether the $5f$ electrons are localized or itinerant. This is beyond the scope of the interpretation of the present experiments, which will be sensitive to the projected electron density. Recall that these experiments are not sensitive to the $5f$ electronic structure below $E_F$, but are sensitive to the electronic structure of the unoccupied states above $E_F$. Many other experiments, notably angular-resolved photoemission, which have observed considerable dispersion of the $5f$ states near $E_F$ \cite{Mydosh,Kawasaki,Durakiewicz,Meng} for \urs\,are consistent with the $5f$ states being itinerant. This is also suggested by the lack of any sharp crystal-field transitions observed in neutron inelastic scattering \cite{Park,Butch}. The majority of theoretical studies have predicted that the $5f$ states are itinerant \cite{Oppeneer}.

\subsection{Valence-band RIXS data}

We also report valence-band RIXS data from three of the compounds with a resolution of $\sim1$~eV. To our knowledge these have not been reported previously from U intermetallics at the U $M_{4,5}$ edges. They show two transitions: (1) the U $P_3$ transition, in which we measure the energy between $6p$ states and the unoccupied $5f$, at between $18$ and $20$~eV and (2) transitions between the valence band states to the unoccupied U $5f$ states. Differences are observed between the values of these transitions for \urs, UPd$_3$, and UO$_2$ – see Fig.~\ref{Fig:7}.

To obtain an idea how these values are related to theory we have performed calculations using the FEFF program to determine the DOSs for the various electron states near $E_F$ in these compounds. The FEFF program has some difficulty with intermetallic materials in locating $E_F$, but it does reproduce the values of the transitions (Fig.~\ref{Fig:8}), which is what is measured in the experiments. These values are also in reasonable agreement with more advanced band-structure determinations of the materials.


\begin{acknowledgements}
We thank the following for samples used in this study. Dai Aoki (CEA-Grenoble) for the \urs, Keith McEwen (UCL, London) for the UPd$_3$, and Philippe Martin (CEA-Cadarache) for the UO$_2$. We thank Patrick Colomp of the Radioprotection services for his help at ID26.\end{acknowledgements}

\bibliography{urs}

\end{document}